# An entropy-based measurement for understanding origin-destination trip distributions: a case study of New York City taxis

Authors: Yuqin Jiang[1], Yihong Yuan[1], Su Yeon Han[1*]
*Corresponding author: su.han@txstate.edu
[1]Department of Geography and Environmental Studies, Texas State University, San Marcos, TX, USA

**Abstract**
A comprehensive understanding of human mobility patterns in urban areas is essential for urban development and transportation planning. In this study, we create entropy-based measurements to capture the geographical distribution diversity of trip origins and destinations. Specifically, we develop origin-entropy and destination-entropy based on taxi and ride-sharing trip records. The origin-entropy for a given zone accounts for all the trips that originate from this zone and calculates the level of geographical distribution diversity of these trips' destinations. Likewise, the destination-entropy for a given zone considers all the trips that end in this zone and calculates the level of geographical distribution diversity of these trips' origins. Furthermore, we have created an interactive geovisualization that enables researchers to delve into and juxtapose the spatial and temporal dynamics of origin and destination entropy, in conjunction with trip counts for both origins and destinations. Results indicate that entropy-based measurements effectively capture shifts in the diversity of trips' geographical origins and destinations, reflecting changes in travel decisions due to major events like the COVID-19 pandemic. These measurements, alongside trip counts, offer a more comprehensive understanding of urban human flows.

## 1. Introduction

Human movements are pulses of the city that define and shape the characteristics of the city. Observing, modeling, and analyzing human movements provide valuable insights into how humans interact with the surrounding built environment (Barbosa et al., 2018; Liu et al., 2015). Understanding human flows in the city is essential for both the city's residents and its developers in multiple perspectives, including infrastructure (Pan et al., 2013; Zhao, Shaw et al. 2016), urban planning and developing (Gonzalez et al., 2008; Wang et al., 2019), business marketing decisions (Yuan et al., 2012; Liu et al., 2017), tourism (Yang et al., 2023; Hu et al., 2019), and emergency management (Jiang et al., 2021; Haraguchi et al., 2022).

Research has leveraged taxi travel data to gain insights into human movement patterns and travel behaviors (Gong et al., 2016; Chen et al., 2018; Peng et al., 2012; Guo et al., 2012), while also enhancing situational awareness within urban environments (Ganti et al., 2013; Kong et al., 2017). Travel behavior refers to the factors related to how and why individuals move from one place to another, including modes of transportation, route and destination choices, frequency and duration of travel, and the purpose of travel (Lanzendord 2002; Van Wee et al., 2013). In the context of travel behavior research, situational awareness involves comprehending traffic dynamics, environmental conditions, and other influential factors on movement and travel decisions, enabling prediction and response to ensure safety, efficiency, and effectiveness in travel and transportation systems (Naseer et al., 2022; Huang & Xiao, 2015). The location information associated with massive amount of vehicles on the road provides researchers unprecedented opportunities to study travel behavior in the city. Mining this information reveals road conditions, people's travel demands, and transportation mode choices.

Traveling with taxi, ride-sharing applications (Uber or Lyft), or charter services is a common transportation mode in densely populated urban areas with low car ownership across the world (Hall et al., 2018; Young & Farber, 2019). The travel records associated with these services provide valuable and rich information for human flows in the city.

Entropy, a concept rooted in physics and information theory, measures the randomness and unpredictability of a system (Shannon, 1948). Geography and spatial science have brought in the concept of entropy to analyze the spatial distribution of geographical features (Batty, 1974; Batty & Mackie, 1972; Batty et al., 2014). If a geographical feature is evenly distributed across the study area, it reaches maximum-entropy. Therefore, entropy is used as a measurement for spatial (in)equity in land use (Rodriguez et al., 2009; Dewa et al., 2022), urban development (Yeh & Li, 2001; Parvinnezhad et al., 2021; Cabral et al., 2013), socioeconomic status (Germano, 2022; Lenormand et al., 2020), and crime (Zhang et al., 2021; Kim & Lee, 2023).

This study advances human mobility studies through its innovative use of an entropy-based measurement approach, applied to taxi data from New York City (NYC) spanning 2016 to 2021. This method sets itself apart from prior studies by employing entropy measurements to quantify the geographical distribution diversity of trip origins and destinations at the taxi zone level. Specifically, we compute origin-entropy and destination-entropy for each taxi zone and explore their temporal changes. The origin-entropy for a taxi zone accounts for all the trips originating in this zone and calculates the geographical distribution diversity of their destinations. Conversely, the destination-entropy for a taxi zone accounts for all the trips that end in this zone and calculates the geographical distribution diversity of their origins. If a taxi zone has a high origin-entropy, it means that trips originated from this zone go to a large number of different destination zones across the city, implying people using taxis to travel to diverse destinations from this zone. A low origin-entropy for a taxi zone means that people who start taxi trips from this zone are likely to select geographically similar destinations (i.e., the trips end in a small range of zones). On the other hand, a taxi zone with a high destination-entropy means that trips ending in this zone come from a large number of different origin zones across the city, indicating that people use taxis to arrive in this zone from diverse origins. A low destination-entropy for a taxi zone indicates that people come from a small range of origins to this destination zone. These entropy-based measurements developed in this study can serve as a valuable adjunct to the conventional measurement of human mobility, which quantifies the number of trips people make between their origins and destinations. Moreover, we have created an interactive geovisualization that empowers users to delve into the spatial and temporal dynamics of origin-destination entropy and travel counts. This is accomplished by leveraging the CyberGIS-Vis, which provides synchronized and coordinated maps and charts that enable changes in one view to be reflected instantaneously in the other (see section 4.4 for the details of CyberGIS-Vis). The insights gleaned from utilizing this tool can provide a distinctive perspective for urban planning and transportation management.

The primary research questions guiding this study are as follows:

1. How can entropy measurement be applied to quantify the geographical diversity of taxi trips in NYC?
2. How can the measurement of entropy unveil novel insights that counting the number of trips might not capture?
3. What temporal variations are identified in both origin entropy, destination entropy, and the quantity of trips for both origins and destinations? How do these temporal changes differ among various taxi zones in NYC, and what insights into travel behaviors can be gleaned from these distinctions?
4. How can the measurement of entropy correspond to major events in NYC, such as transportation-related policy changes and the COVID-19 pandemic?

The remainder of this paper is organized as follows. Section 2 reviews related studies, providing background in human mobility and entropy research. Section 3 introduces the study area and datasets. Section 4 defines essential terms and describes the methodology about the entropy-based measurements as well as an interactive geovisaulization we created to visualize the output. Section 5 presents the results and zone-of-interest exploration. Finally, section 6 discusses future research directions and concludes this study.

**2. Background**
2.1. Human mobility using taxi data.
Research leveraging taxi travel records has been conducted from four distinct perspectives. The first is trip classification and clustering analysis. These studies employed classification and clustering algorithms to identify the hotspots and the dominant travel patterns in the urban environment. By examining the spatiotemporal patterns of hotspots and trip flows, city planners can achieve better understandings about how humans interact with their surrounding physical environments. Insights into travel hotspots and dominant flows help local business to optimize their location selections and urban planners to improve infrastructure for higher transportation efficiency (Guo et al., 2020; Yao et al., 2018; Zhu & Guo, 2014). The second direction is to analyze travel purposes and land use in the city. Studies in this direction examine taxi travel demand associated with different land use types to advance understandings in human-environment interactions in the urban environment (Pan et al., 2012; Liu et al., 2012; Choi et al., 2022; Liu et al., 2015; Qian et al., 2015). The third research direction focuses on event and anomaly detection using massive taxi travel records. These studies apply statistical and machine learning methods to identify anomalies or events such as marathons and parades in the study area (Jiang et al., 2022; Zhu & Guo, 2017; Zhang et al., 2015). Lastly, taxi trips can reveal road conditions. Studies in this direction rely on real-time information provided by taxicabs, such as location and real-time driving speed, to estimate congestion situations. Utilizing real-time traffic data from taxicabs, these studies generated data-driven road condition monitoring systems and can help transportation planners to optimize infrastructure usage in the city (An et al., 2018; Keler et al., 2017; Jiang et al., 2021; Fu et al., 2022; Tang et al., 2015).

2.2. Entropy measurements in human mobility
Entropy, a concept rooted in physics and information theory, measures the randomness and unpredictability of a system (Shannon, 1948). Geography and spatial science have brought in the concept of entropy to analyze the spatial distribution of phenomena (Batty, 1974; Batty & Mackie, 1972; Batty et al., 2014). Since then, entropy has been used in various fields and applications, such as measuring the diversity of land use land cover (Rodriguez et al., 2009; Dewa et al., 2022) and urban sprawl (Yeh & Li, 2001; Parvinnezhad et al., 2021; Cabral et al., 2013).

Entropy, in the context of human mobility at the individual-level, quantifies the degree of randomness or unpredictability in an individual's movements. This measurement assigns probabilities to each distinct location visited by an individual. These probabilities are derived from the frequency or occurrence of visits to each place relative to the total number of locations visited. The entropy captures these probabilities, emphasizing the significance of rare or less frequently visited places, which ultimately quantifies the uncertainty or unpredictability in an individual's movements. A higher entropy value indicates a more diverse and random mobility pattern, while a lower value indicates a more ordered and predictable sequence of visited locations (Song et al., 2010; Gonzalez et al., 2008; Qin et al., 2012; Zhao et al., 2019).

Various studies have employed entropy as a metric to measures the diversity and unpredictability of an individual's activities based on the locations they visit and the frequency of those visits at the individual level (Zhao, Tarkoma et al. 2016; Xu et al., 2018). More specifically, entropy was utilized as an indicator to analyze individual's activity space geotagged social media posts and photos. By calculating the entropy based on individual's movements, the researchers explored the social media users' travel interests (Wang and Yuan, 2021; Yuan and Medel, 2016; Yuan and Raubal, 2012; Yuan and Wang, 2018). Similar measures have also been applied to cell phone-based movements. For instance, Yuan et al. (2012) utilized entropy to measure the regularity of visitation patterns exhibited by individual cell phone users.

While entropy is traditionally applied to measure the diversity and unpredictability of individual activities, its utility extends to the community level, offering profound insights into a wide range of phenomena that surpass simple movement patterns of individuals. For example, Huang et al. (2022) applied entropy to gauge the disorder in demographic and socio-economic characteristics among communities linked by frequent human spatial interactions. Moncure (2017) utilized an entropy-based index to assess accessibility to healthy food options in different neighborhoods. Additionally, Ning et al. (2022) combined entropy with other metrics to enhance equity in community-level facility access for the elderly. Furthermore, Zhang et al. (2021) and Kim & Lee (2023) employed an entropy-based index to assess the diversity of points of interest (POIs) within specific aggregated places, such as census block groups or at a 250-meter grid level.

While entropy measurements at the community level have yielded significant insights in areas beyond human mobility, their application in understanding the geographical distribution diversity of trips at the community level remains underexplored. This research addresses this gap by applying entropy to quantify the diversity of geographical distributions of taxi trips in NYC at taxi zone (i.e. community) levels, rather than on an individual basis. Our study aims to illustrate how measuring community-level entropy, combined with straightforward counts of trips at origins and destinations, can enhance our understanding of human mobility patterns. Furthermore, this approach generates valuable insights that can guide urban and transportation planning, event detection, and policymaking, emphasizing the practical benefits of applying group-level entropy in these fields.

3. **Data and study area**

3.1. Datasets

Taxi dataset was provided by New York City Taxi and Limousine Commission (NYC TLC). NYC TLC operates the taxicabs in the city and regulates licenses for ride sharing services (NYC TLC, n.d.). Therefore, the datasets received from NYC TLC included a wide range of services, including the traditional taxicabs, main ride sharing applications (Uber and Lyft), and some small local ride sharing service providers. In this paper, we do not distinguish these taxi types and thus, unless specified, the term "taxis" refers to both the traditional taxis and ridesharing vehicles.

All vehicles registered with NYC TLC are mandated to use GPS tracking for their trip recording. Therefore, this dataset includes all the taxi and ride-sharing trips in NYC, with occasional technical issues resulting in some trips not being recorded correctly. As a result, there are no spatiotemporal collection or sampling biases involved in the process. Specifically, the dataset spans from January 1st, 2016, to July 31st, 2021, capturing a total of 1,304,353,323 taxi and ride-sharing trips. However, the number of vehicles is not provided. Figure 1 shows the trend line for averaged daily trips by month.

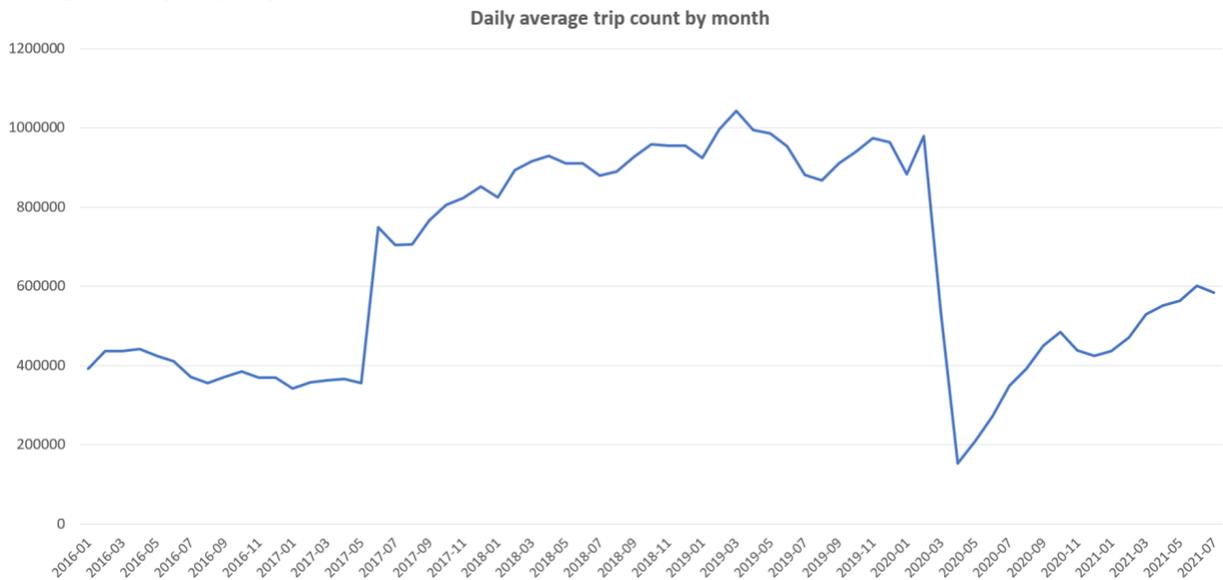

Figure 1. Daily average trip count by month.

To protect privacy, NYC TLC does not provide the exact trip starting and ending locations, instead, they used a taxi zone. There are 263 zones covering all the 5 boroughs in NYC plus Newark airport (EWR) in Newark, New Jersey. Each zone is a group of blocks nearby and is roughly based on NYC Department of City Planning's Neighborhood Tabulation Areas (NYC TLC, n.d.). Some zones are designated for a specific function, such as Central Park and JFK airport; while some other zones cover mixed land functions and may include residential, commercial, and open spaces. Each trip record includes the following information: vendor ID, pick-up date time, drop-off date time, passenger count, trip distance, rate code, pick-up location ID, drop-off location ID, payment type, fare amount, extra charge, MTA tax, tip amount, tolls amount, improvement surcharge, total amount, congestion surcharge.

We first filtered out trip records with missing origin or destination zones. This can be resulted from error in GPS recording. Then, we re-organized all the trip records into zone-to-zone trips daily. After this step, we processed datasets only containing the following information: date of the trip, origin zone, destination zone, and count. In other words, we organized our datasets in a way that counts the number of trips that occurred between a pair of origin-destination zones on each day.

3.2. Study area

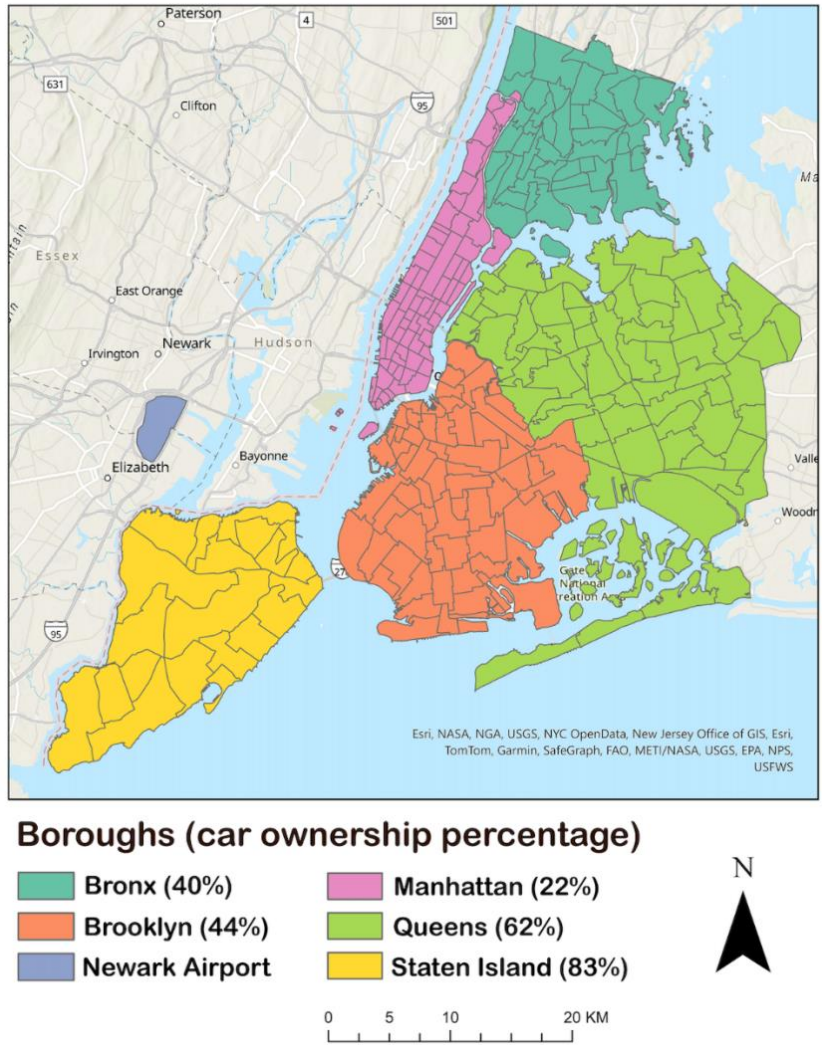

Figure 2. NYC taxi zones in the 5 boroughs and Newark Airport (EWR) in New Jersey. The number in the parentheses denotes the car ownership percentage in each borough.

Figure 2 shows the taxi zones and the 5 boroughs of NYC. New York City (NYC) is the study area for this study. According to the Census, NYC had an estimated population of 8.8 million in 2020 (NYC Planning, n.a.). It has the highest population density in the United States. NYC consists of five boroughs and each borough is a county: Brooklyn (Kings County), Bronx (Bronx County), Manhattan (New York County), Queens (Queens County), and Staten Island (Richmond County).

According to the American Community Survey, only 45% households in NYC owned at least one car in 2019. In addition, car ownership is not evenly distributed across the five boroughs. In Manhattan has the lowest car ownership at 22%, while in State Island is 83%. Due to this low car ownership, many NYC residents, especially in Manhattan, heavily rely on taxis for their daily travel needs. Therefore, taxis play an important role in providing accessible and convenient mobility solutions for residents without a car (NYCEDC, 2018).

4. Methodology

Our study includes three main steps: data processing, measurement calculation, and visualization. We first re-constructed all the trip records into a daily basis. Then, we calculated the daily origin count, destination count, origin-entropy, and destination-entropy for each taxi zone. Lastly, a CyberGIS-based visualization application was created for exploration and analysis.

4.1. Data processing

The original taxi datasets were organized monthly, with each month containing multiple files, one for each taxi type. We restructured the data into a daily basis. For each day, we summarized all the trips by different taxi types. After this step, we created 2,038 separated files, one for each day's travel records. These reorganized datasets made it easier for daily analysis. In the second step, we focused on constructing origin-destination flows. For each day, we counted the number of trips between every unique origin-destinations pair. This analysis resulted in a travel count table for each day, which includes three columns: origin, destination, and count. This aggregated travel summarizations served as the foundation for entropy calculation and visualization.

4.2. Entropy-based measurement calculation

In this study, we innovatively applied the concept of entropy to measure the variety of zone-to-zone trips. The original entropy measurement in information science by Claude Shannon (1948) is defined as shown in Eq. 1.::

$$H = -\sum_{i} p_i \log(p_i)$$

Eq. 1.

Eq. 1. shows the entropy calculation in the discrete case, where $p_i$ is the distribution probability of the unit $i$. It can be interpreted as the level of uncertainty for a distribution. For example, given 4 individuals living in 4 zones.

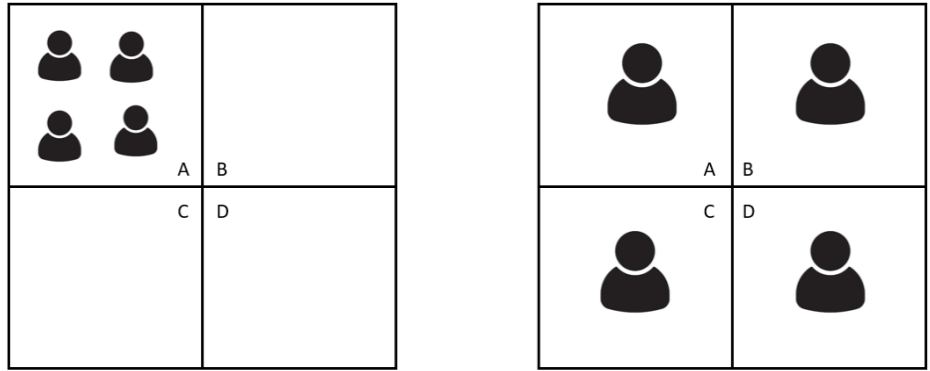

Figure 3. Illustration of a simple entropy calculation

In the first scenario (Figure 3 left), all 4 individuals live in the same zone. These 4 individuals present low geographical distribution diversity, as they are all clustered in one zone. If a person is randomly selected, the probability of this people is picked from zone A is 100%, and 0% for zone B, C, and D. In other words, the uncertainty level in this scenario is 0. Taking this into the entropy equation (Eq. 2.), it can be calculated as following:

$$H = -\sum_i p_i \log(p_i)$$
$$= -(p_A \log p_A + p_B \log p_B + p_C \log p_C + p_D \log p_D)$$
$$= -\left(\frac{4}{4} \log \frac{4}{4}\right)$$
$$= 0$$

Eq. 2.

In another scenario, there is only one person living in each zone (Figure 3 right). These four individuals are spread out across all zones, showing a high diversity in their geographical distribution instead of being clustered in only one or two specific zones. If one person is randomly selected, the probability of this person is picked from zone A is identical to that of zone B, C, and D, which is 25%. It is uncertain from which block a person would be randomly selected. In other words, the uncertainty level in the second scenario is higher than that of the first scenario. The entropy calculation for this scenario is shown as the following Eq. 3.:

$$H = -\sum_i p_i \log(p_i)$$
$$= -(p_A \log p_A + p_B \log p_B + p_C \log p_C + p_D \log p_D)$$
$$= -\left(\frac{1}{4} \log \frac{1}{4} + \frac{1}{4} \log \frac{1}{4} + \frac{1}{4} \log \frac{1}{4} + \frac{1}{4} \log \frac{1}{4}\right)$$
$$= 0.6021$$

Eq. 3.

## 4.3. Entropy-based measurements for origin-destination taxi trips

We have developed measurements based on the concept of entropy to assess the geographical diversity in the distributions of origins and destinations of taxi trips across different zones. The measurements are intended to complement the traditional method of measuring human mobility, typically focused on counting the number of trips made. Specifically, for each taxi zone, four measurements are used to quantify taxi trip patterns: (1) origin count; (2) origin-entropy; (3) destination count; and (4) destination entropy. Following are the definitions:

**Origin count ($N_{O,i,t}$)** is the number of taxi trips starting from taxi zone $i$. It represents that there were $N_{O,i,t}$ trips leaving taxi zone $i$ on day $t$.

**Origin-entropy ($H_{O,i,t}$)** represents the origin-entropy for taxi zone $i$, on day $t$. It accounts for all the taxi trips started from taxi zone $i$, on day $t$, and calculates the geographical distribution diversity of these trips' destinations. In other words, the origin-entropy measures how spread passengers are heading away from their origin zone $i$.

**Destination count ($N_{D,j,t}$)** is the number of taxi trips ended in taxi zone $j$. It represents that there were $N_{D,j,t}$ trips arriving in taxi zone $j$ on day $t$.

**Destination-entropy ($H_{D,j,t}$)** represents the destination-entropy for taxi zone $j$, on day $t$. It accounts for all the taxi trips ended in taxi zone $j$, on day $t$, and calculates the geographical distribution diversity of these trips' origins. In other words, the destination-entropy measures how the spread of the locations from which passengers are coming to their destination zone $j$.

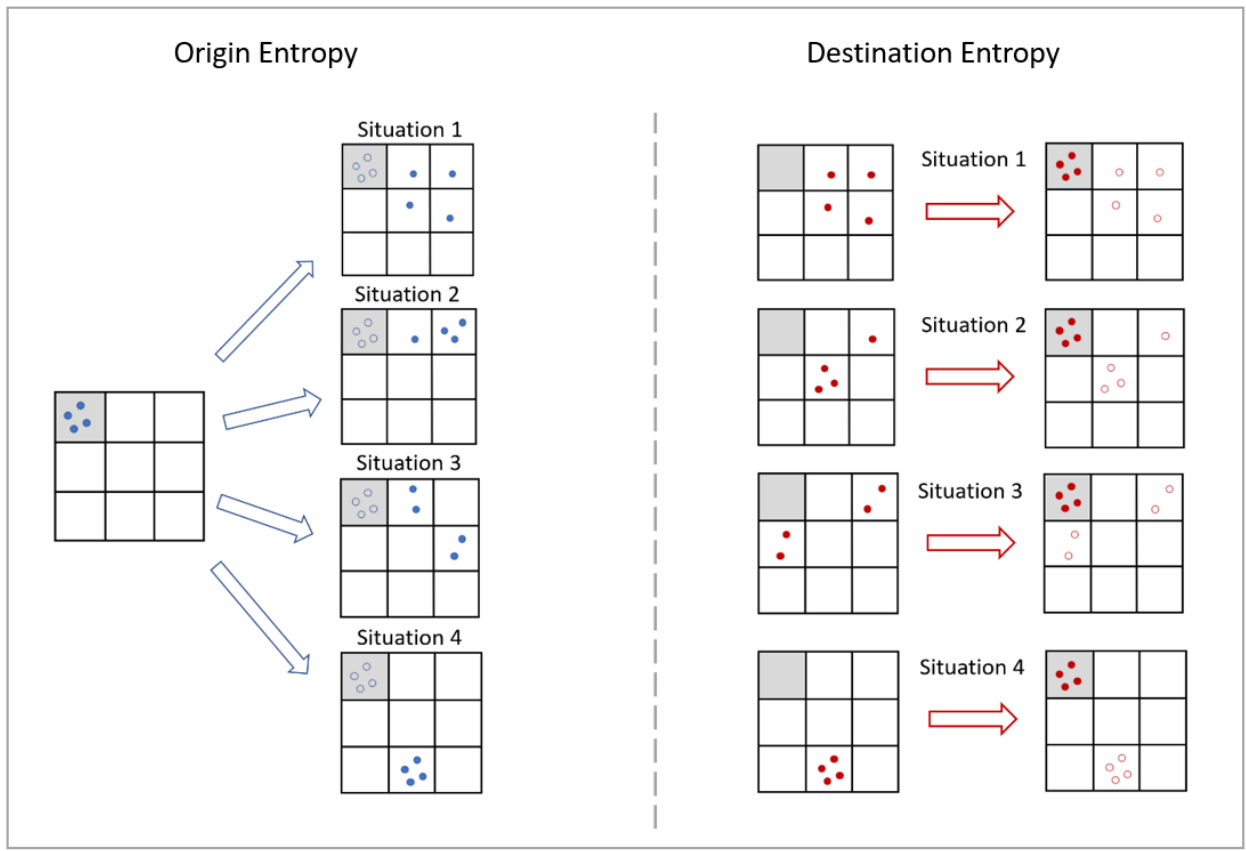

Figure 4. Demonstration of origin entropy (left) and destination entropy (right). Nine zones are represented in each box. The solid circles in the boxes on the left side of the arrows represent the hypothetical origin locations of each trip, while the solid circles in the boxes on the right side of the arrows represent the hypothetical destination locations after each trip. Hollow circles indicate the original locations before the move is made.

The origin-entropy for Zone $i$ is measured by the following Eq. 4:

$$H_{O,i,t} = - \sum_j \frac{n_j}{N_{O,i,t}} \log\left(\frac{n_j}{N_{O,i,t}}\right) \qquad \text{Eq. 4,}$$

Where $H_{O,i,t}$ is the origin-entropy for Zone $i$ on Day $t$. $N_{O,i,t}$ is the total number of trips leaving Zone $i$ on Day $t$. In other words, on Day $t$, Zone $i$ is the origin ($O$) for a total number of $N_{O,i,t}$ trips. $n_j$ is the number of trips ended in destination Zone $j$.

Similarly, the destination-entropy for Zone $j$ is measured by the following Eq. 5:

$$H_{D,j,t} = - \sum_i \frac{n_i}{N_{D,j,t}} \log\left(\frac{n_i}{N_{D,j,t}}\right) \qquad \text{Eq. 5,}$$

where $H_{D,j,t}$ is the destination-entropy for Zone $j$ on Day $t$. $N_{D,j,t}$ is the total number of trips ended in Zone $j$ on Day $t$. In other words, Zone $j$ is the destination ($D$) for a total number of $N_{D,j,t}$ trips. $n_i$ is the number of trips started in origin Zone $i$.

Using Figure 4 as an example, in the left panel, there are four trips starting from the origin zone (the shaded zone). In situation 1, the four trips all ended in different destination zones. In situation 2, three trips ended in one zone, and one trip ended in another zone. In situation 3, two trips ended in one zone and the other two trips ended in another zone. In situation 4, the four trips all ended in the same destination zone. If the origin count — i.e., quantifying the number of trips that originate from a specific zone — is the only measurement used, it would record all the four situations identical, as in all the situations, four trips were observed. However, the origin-entropy can capture the differences. Equations 6 to 9 show the calculation for situations 1 to 4 respectively.

$$\begin{aligned} H_{O,i,t} &= - \sum_j \frac{n_j}{N_{O,i,t}} \log\left(\frac{n_j}{N_{O,i,t}}\right) \\ &= -(p_A \log p_A + p_B \log p_B + p_C \log p_C + p_D \log p_D) \\ &= -\left(\frac{1}{4}\log\frac{1}{4} + \frac{1}{4}\log\frac{1}{4} + \frac{1}{4}\log\frac{1}{4} + \frac{1}{4}\log\frac{1}{4}\right) \\ &= 0.6021 \end{aligned} \qquad \text{Eq. 6 (Situation 1).}$$

$$\begin{aligned} H_{O,i,t} &= - \sum_j \frac{n_j}{N_{O,i,t}} \log\left(\frac{n_j}{N_{O,i,t}}\right) \\ &= -(p_A \log p_A + p_B \log p_B) \\ &= -\left(\frac{1}{4}\log\frac{1}{4} + \frac{3}{4}\log\frac{3}{4}\right) \\ &= 0.244 \end{aligned} \qquad \text{Eq. 7 (Situation 2).}$$

$$\begin{aligned} H_{O,i,t} &= -\sum_j \frac{n_j}{N_{O,i,t}} \log\left(\frac{n_j}{N_{O,i,t}}\right) \\ &= -(p_A \log p_A + p_B \log p_B) \\ &= -\left(\frac{2}{4} \log \frac{2}{4} + \frac{2}{4} \log \frac{2}{4}\right) \\ &= 0.301 \end{aligned}$$

Eq. 8 (Situation 3).

$$\begin{aligned} H_{O,i,t} &= -\sum_j \frac{n_j}{N_{O,i,t}} \log\left(\frac{n_j}{N_{O,i,t}}\right) \\ &= -(p_A \log p_A + p_B \log p_B + p_C \log p_C + p_D \log p_D) \\ &= -\left(\frac{4}{4} \log \frac{4}{4}\right) \\ &= 0 \end{aligned}$$

Eq. 9 (Situation 4).

Based on these calculations, situation 1 has the highest origin-entropy, as the four trips started from the origin zone travel to four different destinations, indicating under this situation the geographical distribution diversity reaches the highest variety level. Situation 4, in which all the four trips end in the same destination, has the smallest origin-entropy, meaning that under this situation the geographical distribution diversity of destinations has the lowest variety level.

Figure 4 right panel illustrates the destination-entropy calculation. In all four situations, there are four trips that end in the same destination zone (the shaded zone). If counting the number of destination trips is the only measurement used, all four situations would be recorded with no difference. Destination-entropy can be calculated using similar equations as shown in origin-entropy (Eq. 6 to 9). The results are: 0.6021, 0.266, 0.301, and 0, for each of the four situations, respectively. Therefore, the destination-entropy measures the geographical distribution diversity level of origins. In summary, as demonstrated above, our origin and destination entropy measurement complement the conventional approach of assessing human mobility, which traditionally emphasizes the tallying of trip counts.

Taking a specific zone as the origin, the origin-entropy quantifies the geographical distribution diversity level of destinations to which people are traveling. Using JFK Airport as an example, the origin-entropy takes all the trips starting from JFK Airport into consideration and measures how spread out those trips' destinations are. If JFK Airport has a low origin-entropy, it means that trips originating from JFK Airport predominantly travel to a small number of different zones. In other words, the diversity in the geographical distribution of these trip destinations is low. Conversely, a high origin-entropy at JFK Airport indicates that trips originating from JFK Airport head to a diverse range of zones, signifying a higher geographical distribution diversity of their destinations. Opposite to the origin-entropy, the destination-entropy considers all the trips ending in a given zone and measures the diversity in the geographical distribution of taxi trip origins. Also using JFK Airport as an example, we consider all the trips ending at the JFK

Airport and calculate the geographical distribution diversity level of origins people are traveling from. If JFK Airport has a low destination-entropy, it means that people arriving at JFK Airport predominantly come from a single or less varied zones. In other words, the geographical distribution diversity of these trips' origins is low. Conversely, a high destination-entropy for JFK Airport indicates that travelers come from variously different zones. In other words, the geographical distribution diversity of their origins is high.

4.4. Cyber-GIS visualization

We employed CyberGIS-Vis to examine spatiotemporal trends in taxi trip origins, destinations, and their entropy. This open-source tool enables interactive geospatial visualization and scalable analytics (*CyberGIS-Vis*, 2021/2023). CyberGIS-Vis facilitates the creation of web-based visualizations with Coordinated Multiple Views (CMV). In CMV, diverse data representations, such as maps and charts, are dynamically interlinked (Roberts, 2007). Thus, modifications in one visualization are instantly reflected in others through operations like cross-filtering, brushing, highlighting, and selection. CyberGIS-Vis offers a range of modules for exploring relationships in geospatial attributes and trends in spatiotemporal datasets.

A key component of CyberGIS-Vis, the Adaptive Choropleth Mapping Mapper (ACM), provides essential features for advanced choropleth mapping (Han et al., 2019). These include (1) automated generation of consistent class intervals for color-coding various choropleth maps, (2) dynamic visualization of local variation in a variable, and (3) synchronized examination of multiple choropleth maps through linking. CyberGIS-Vis users can integrate ACM with various charts such as stacked charts, correlograms, scatter charts, parallel coordinate charts, and comparison line charts to create Coordinated Multiple Views. We specifically utilized two modules aimed at spatiotemporal data visualization: the Multiple Linked Chart (MLC) and the Comparison Line Chart (CLC).

CyberGIS-Vis offers a suite of visualization modules that are reproducible and adaptable, enabling users to easily tailor visualizations like MLC and CLC by adjusting input parameters and substituting their own data into provided example code. Enhancing the reproducibility of CyberGIS-Vis, its integration into the CyberGISX environment enables users to create interactive geovisualizations without the need for installing any additional software or libraries. CyberGISX is an online platform designed for advanced geospatial analysis, combining cyberinfrastructure and GIS technologies within a user-friendly JupyterLab environment (Kang et al., 2019; CyberGISX, n.d.). Specifically, our process of creating visualizations with CyberGIS-Vis began with the sample Python code from the CyberGISX website, which illustrates the use of the platform for visualization (Han et al., n.d.). By starting with templates like 'Adaptive Choropleth Mapper with Multiple Linked Chart (MLC)' and 'Adaptive Choropleth Mapper with Comparison Line Chart (CLC)', we input our own data sets, tweaked the input parameters, and executed the 'Adaptive_choropleth_Mapper_viz' function to generate our custom visualizations. The parameter descriptions of CyberGIS-Vis is available in the section of input parameters at their GitHub site (*CyberGIS-Vis*, 2021/2023). The reader is referred to the Appendix to see the details of the input parameters we set and how to input our datasets.

5. **Results and discussion**

In the results section, we present visualizations for four measurements: origin count, destination count, origin-entropy, and destination-entropy. The details of origin-entropy, and destination-entropy are described in section 4.3. The origin count and destination count are the number of trips that started and ended, respectively, in a particular zone. They are the most commonly used indicators to measure human mobility in existing studies. By utilizing these four measurements, our objective is to identify significant events, such as policy changes or major city-wide events, that alter taxi-related travel behaviors, reflecting preferences in traveling with taxi.

5.1. Interactive visualization of taxi trips and entropy using CyberGIS-Vis
5.1.1. MLC for NYC taxi trips and entropy

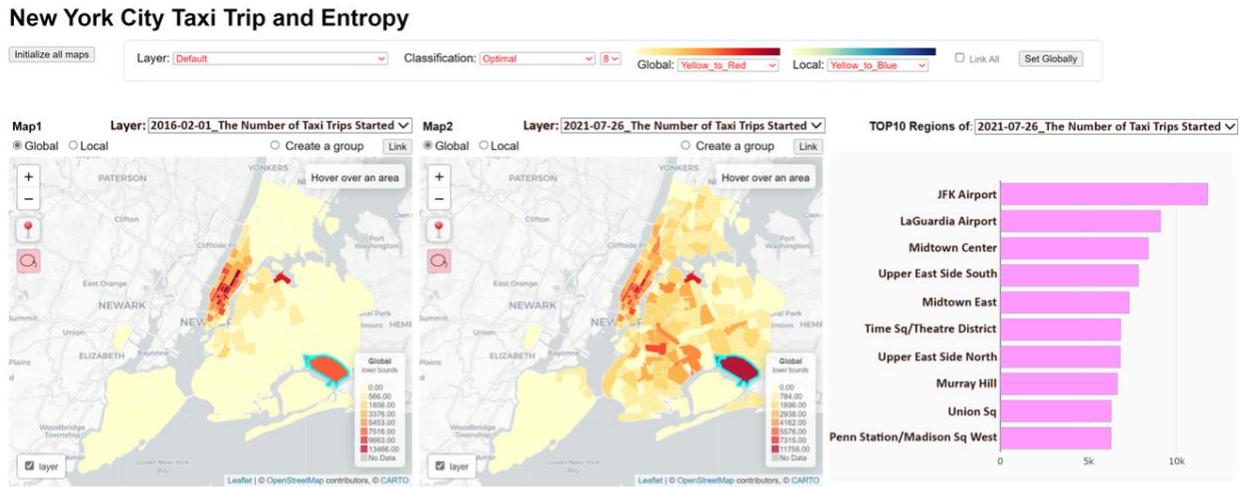

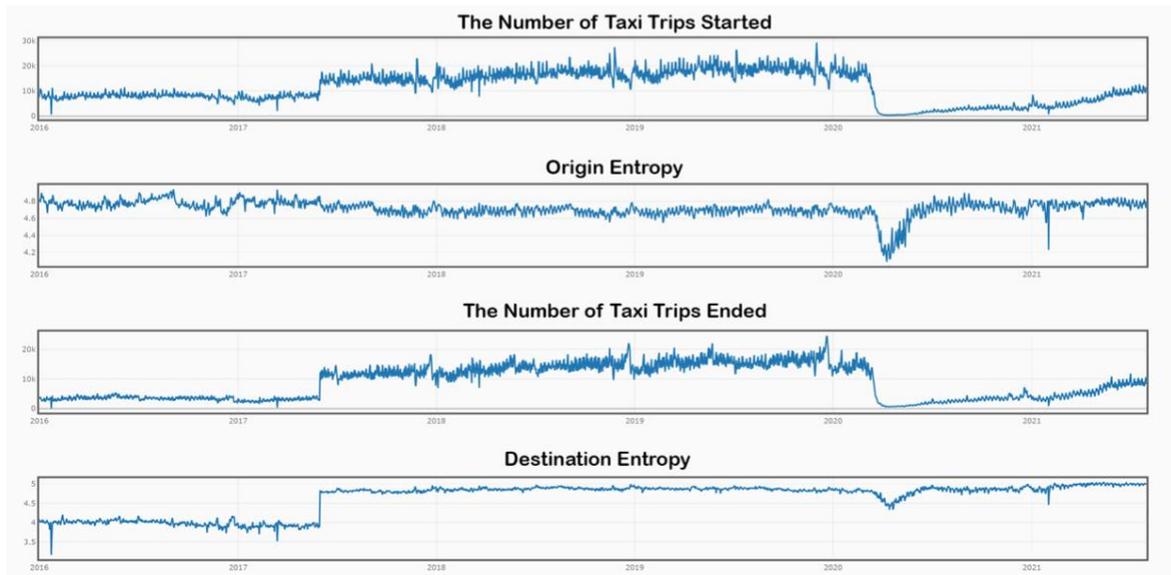

Figure 5. A screenshot of MLC visualization page with JFK Airport selected to display the temporal trend lines. A video demonstrating how to navigate data using this view is available at: https://su-gis.org/mobility/Figure5

The first visualization application is MLC, which allows users to select a zone of interest and examine the daily changes in the number of trips and entropy for one zone from 2016-01-01 to 2021-07-31. An interactive visualization is available at https://su-gis.org/mobility/MLC. Figure 5 shows one of visualizations created from this application. The upper part includes two maps and a bar chart. The two maps provide a comparison between different measurements or time periods. For the map comparison, we set the same class intervals on both maps by checking "Link All" at the top. Users can use the dropdown list to select the date and content (trips or entropy) to be shown in the comparison. The right chart on the upper panel shows the top 10 regions of the selected contents in Map 2. The lower panel shows the trend line measurements for Origin Count, Origin-Entropy, Destination Count, Destination-Entropy. In the example shown in Figure 5, we selected JFK Airport as our zone of interest.

### 5.1.2. CLC for NYC taxi trip and entropy comparison

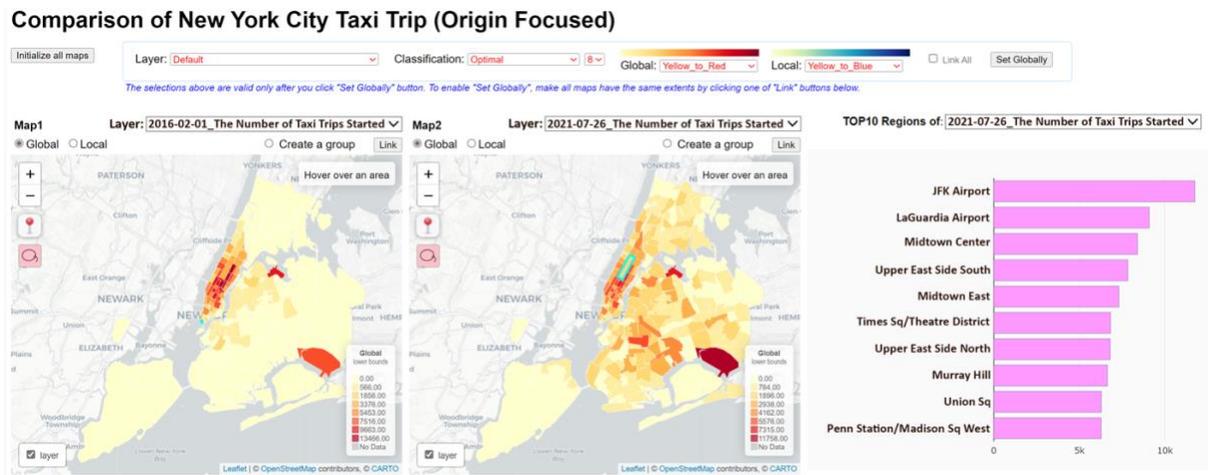

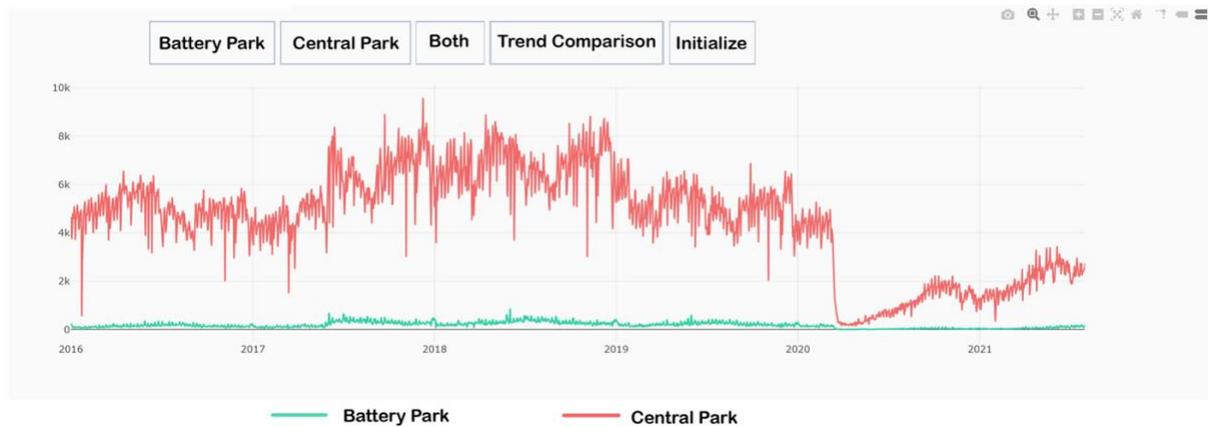

Figure 6. CLC visualization page. A video demonstrating how to navigate data using this view is available at: https://su-gis.org/mobility/Figure6.

CLC is the second visualization application, which allows users to choose two zones of interest and explore changes from 2016-01-01 to 2021-07-31. We created four different visualization pages, each corresponding to one of the four measurements. These pages can be accessed at https://su-gis.org/Mobility/CLC. The example shown in Figure 6 is CLC for Origin Count, which provides the trend lines for the number of trips started from two taxi zones. Same as the MLC visualization page, the upper part includes two maps, providing spatial distribution comparison for two different days. There's also a bar chart showing the top ten regions with the highest number on the selected day, which is the day selected for map 2 (the right map). In Figure 6, we selected Central Park and Battery Park for comparison. In the lower panel, there are five buttons: names for the two selected locations, *Both, Trend Comparison,* and *Initialize*. The functions of each button are detailed in the following paragraph.

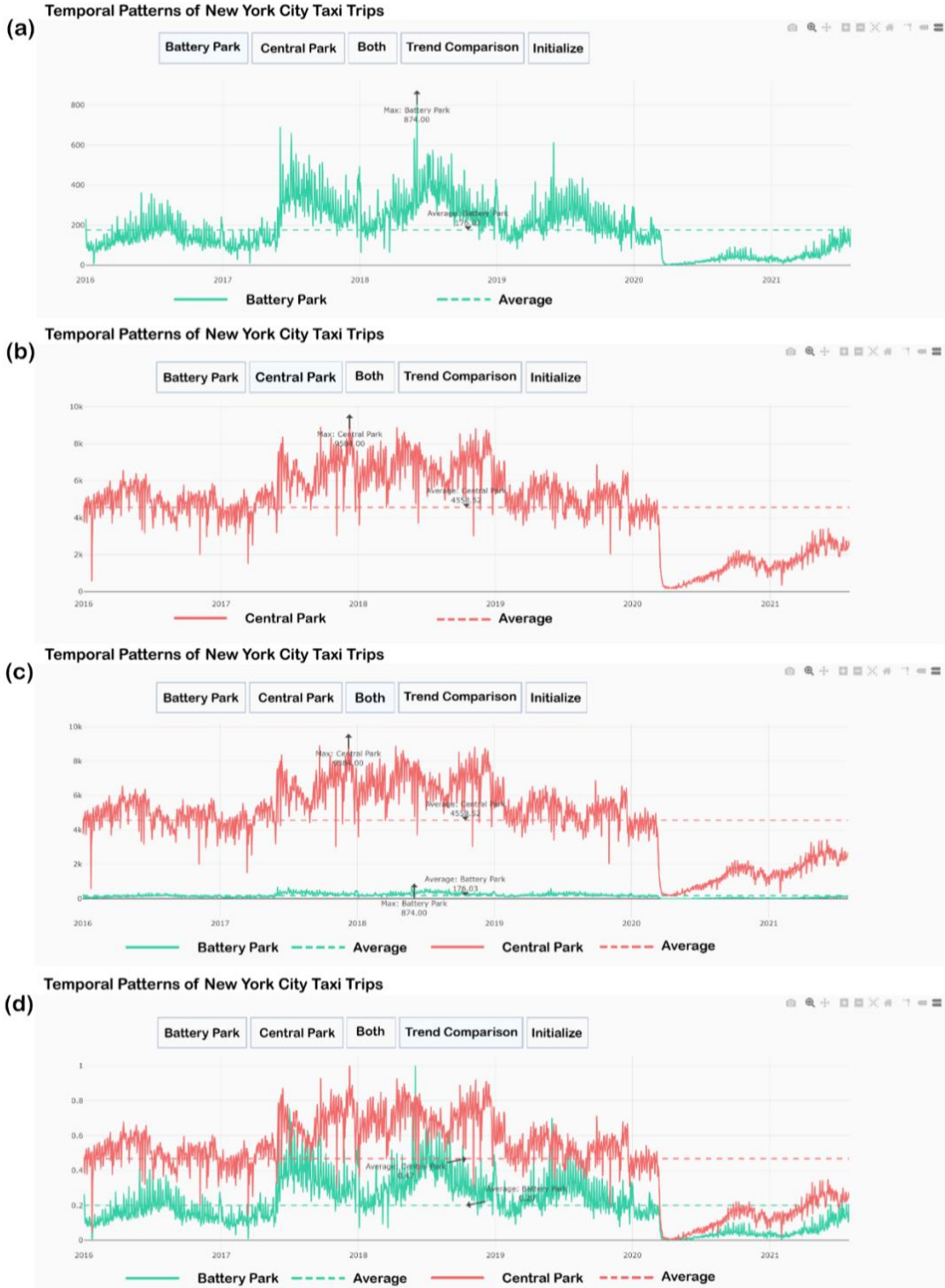

Figure 7. Trend lines comparison with different functions. (a) When *Central Park* button is activated, it only shows the temporal trend line for Central Park. (b) When *Battery Park* button is activated, it only shows the temporal trend line for Battery Park. (c) The function *Both* displays temporal trend lines for both zones together and provides dash lines for average value. (d) *Trend Comparison* normalizes both trend lines into a range between 0 and 1. These images were partially obtained through screen captures as we explored the data, creating several different views starting from those shown in Figure 6.

Figure 7 (a)-(d) demonstrate the trend lines visualization corresponding to each of the four buttons upon their selection. The first and second button are the names for the two selected locations. Figure 7 (a) and (b) show the temporal trend lines for Central Park (green line) and Battery Park (red line) separately when the corresponding button is activated. The third button, labeled 'Both', overlays the temporal trend lines for both locations onto a single chart, which provides a direct comparison of the temporal patterns between these two locations (Figure 7 (c)). The fourth button, labeled 'Trend Comparison' standardizes both trend lines within a 0 to 1 range, allowing for an equitable comparison of their temporal fluctuations irrespective of their original data ranges (Figure 7 (d)). This function captures the seasonal and periodical fluctuations. The last button, 'Initialize', returns to the original visualization of the trend lines.

5.2.  Zone-of-interest exploratory analysis

We investigated the varying patterns of taxi travel across distinct areas in NYC. Some zones serve unique functions such as transportation hubs and recreational spaces and some zones are a group of nearby blocks that contain mixed land uses. In the MLC demonstrations, we chose JFK airport to exemplify the international transportation hub, Central Park and Flushing Meadow-Corona Park for outdoor recreational activities. In the CLC illustrations, we compared Central Park and Battery Park to discern the different characteristics between these two green spaces.

5.2.1. MLC-based patterns

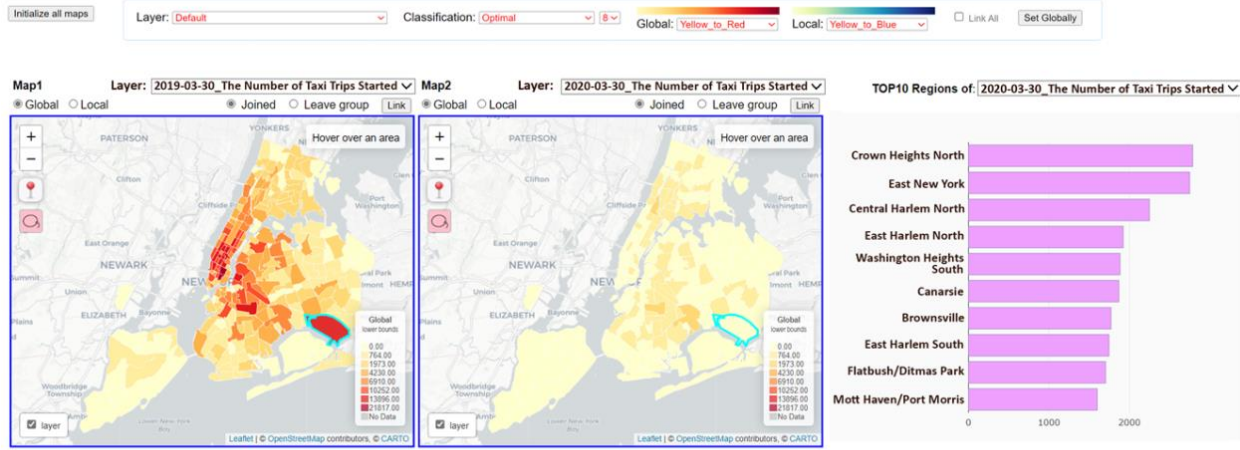

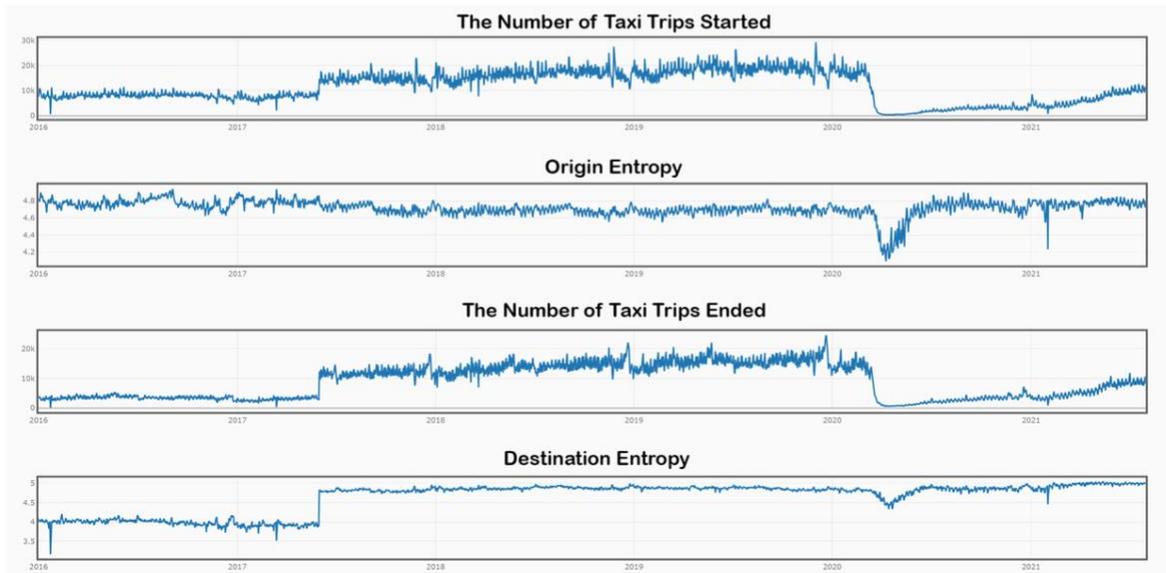

Figure 8. MLC visualization for JFK Airport.

In the map comparison in Figure 8, we selected Origin Count for 2019-03-30 (left map) and 2020-03-30 (right map) to examine the impact of the COVID-19 pandemic on taxi trips. A deeper red presents a larger number of trips originated from this taxi zone. Comparing these two maps, the left map for 2019-03-30 displays a deeper red color, while those zones in the right map for 2020-03-30 appeared lighter. The contrast between these two maps offers obvious visual proof of a substantial decrease in taxi activities during the COVID-19 pandemic.

The four trend lines show Origin Count (the number of taxi trips started), Origin-Entropy, Destination Count (the number of taxi trips ended), and Destination-Entropy, respectively. Noticeably, in late June 2017, there's a substantial rise in both origin count and destination count. This increase can be attributed to the state-level approval for ride-sharing services, such as Uber and Lyft. Although ride-sharing services were already permitted within the city limit of NYC, extending this permission to the state-level significantly increased the number of drivers operating in NYC. The impact of this announcement is reflected in the trend lines for origin count and destination count in many taxi zones. For JFK airport, we observed the increases in origin count, destination count, and destination-entropy, but the origin-entropy did not have substantial changes in 2017. This reflects the different travel behaviors between taxi trips started from and ended in JFK airport. Passengers leaving JFK airport using taxis were heading to various destinations, which was indicated by the consistently high origin-entropy. However, those seeking transportation to the airport often faced challenges in finding empty taxicabs on the street or arranging a ride with limited number of ride-sharing drivers. As a result, they may choose alternative transportation modes, such as subways, bus, or arrange rides with private services. This explains why those who arrived at JFK airport in taxi were more likely to come from a limited amount of origin zones, where were favored by vacant taxi drivers to seek passengers. After the state-level ride-sharing legalization, there was a surge in the number of drivers, making it more convenient for people to arrange a ride from the areas where finding a vacant taxicab was challenging before. As a result, there was a rise in the destination-entropy of JFK Airport, as more passengers took taxis from a larger geographical distribution diversity of origin zones.

In March 2020 when the COVID-19 pandemic hit, we observed drops in all four measurements. Due to the lockdown policy and the reduction of airplane travel, both origin count and destination count decreased significantly. Although the counts had slowly increased since 2020 summer, by 2021 summer, they remained significantly lower than the pre-pandemic level. On the other hand, both origin-entropy and destination-entropy present different trends than the counts. Although both origin- and destination-entropy decreased in March 2020 when the pandemic just started, by 2020 summer, both entropy measurements already recovered to the pre-pandemic level. These entropy measurements indicate that people who took taxis to and from JFK Airport were still heading to a large variety of destinations and coming from diverse origins, despite the reduction in the number of trips.

The origin-entropy had a significant decrease in early 2021, which corresponds to a snowstorm in NYC area between 2021-01-31 and 2021-02-02 (National Weather Service, 2021). This storm caused massive flight cancellations and ground transportation service suspensions. Due to this reason, taxis left JFK airport were heading towards a limited amount of destination zones, such as nearby hotels. Although we found minor drops in both origin count and destination count, they were not as obvious as the drop for origin entropy. These decreases can be caused by both reduced number of passengers who wanted to leave the airport and less drivers willing to drive during the storm.

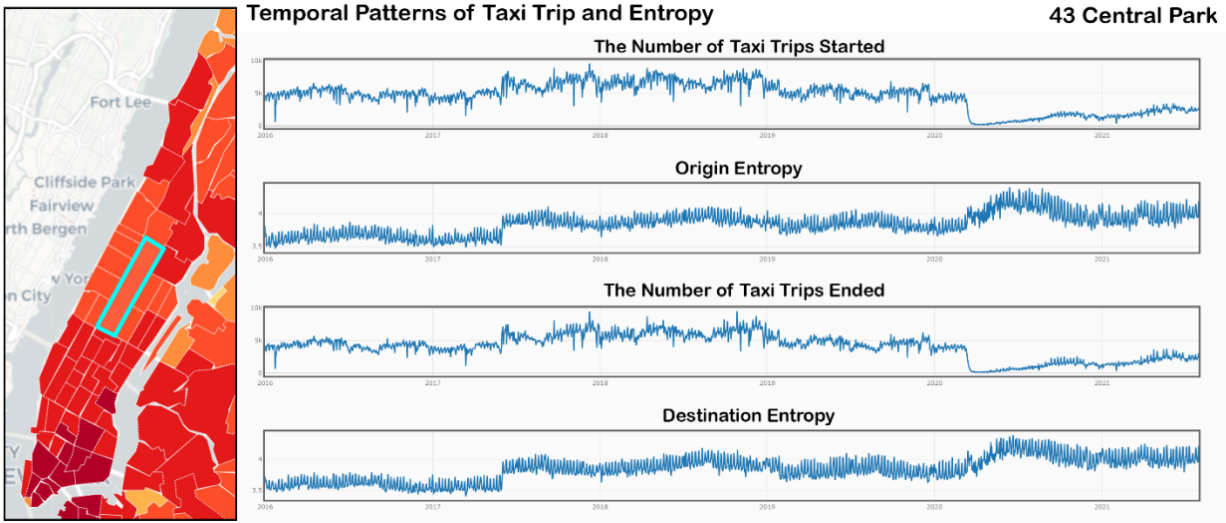

Figure 9. Trend lines for the four measurements for Central Park. Its location is highlighted on the left map. These images were partially obtained through screen captures as we explored the data, creating several different views starting from those shown in Figure 5.

Figure 9 shows the trend lines for the four measurements for Central Park. Central Park, located in the heart of Manhattan and surrounded by skyscrapers, is one of the most well-known public parks in the world. Similar to JFK Airport, the origin count and destination count for Central Park increased since June 2017, indicating that extending ride-sharing services at the state level boosted taxi activities. At the same time, both origin-entropy and destination-entropy also increased, indicating that the ride-sharing services expanded the geographical distribution diversity level of passengers' origins and destinations. In March 2020, both origin count and destination count had a sharp decline due to the strict lockdown policies in NYC. Paradoxically, both origin- and destination-entropy increased during the pandemic. Central Park was a popular destination for outdoor activities. It is easily accessible with multiple subways and buses nearby. However, after the pandemic, concerns were raised about getting infected in densely populated public transit. Consequently, a growing number of New Yorkers chose to use taxis instead. While the number of trips started from and ended in Central Park substantially decreased compared to pre-pandemic level, these taxi trips presented larger geographical distribution diversity level in both origins and destinations. This change could have resulted from people living in well-connected zones with public transit switching to use taxi to avoid possible infections during the pandemic, and therefore, increased the origin- and destination-entropy.

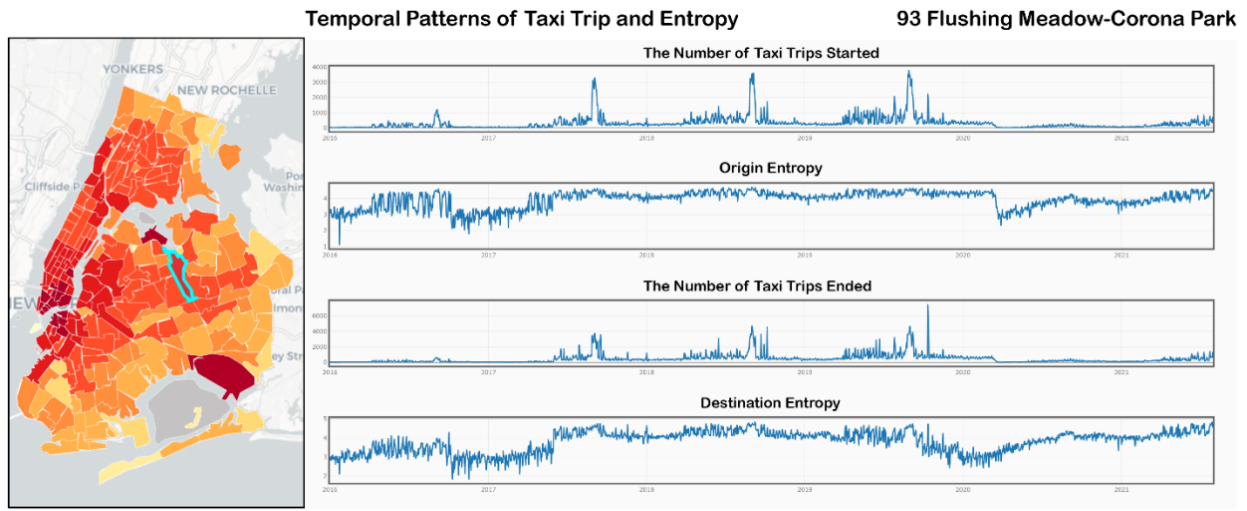

Figure 10. Trend lines for the four measurements for Flushing Meadows-Corona Park. Its location is highlighted on the left map. These images were partially obtained through screen captures as we explored the data, creating several different views starting from those shown in Figure 5.

Figure 10 illustrates the trend lines for Flushing Meadows-Corona Park. This taxi zone includes one of the most iconic parks in Queens, NY, a popular destination for daily outdoor activities. Additionally, it also includes Citi Field, the home stadium for Major League Baseball (MLB) team New York Mets, and Arthur Ashe Stadium, the main stadium for U.S. Open tennis tournament. This taxi zone presents obvious seasonal patterns in both origin and destination counts. Notably, during the U.S. Open, we observed sharp increases in the number of trips started from and ended in this zone, indicating a larger number of spectators used taxi as their primary transportation mode. On 2019-10-12 and 2019-10-13, a music festival was held in Citi Field, which attracted over 60,000 attendees (NY1 NEWS, 2019). The destination count on these two days was recorded highest throughout our data span, indicating a lot of audience arrived at Citi Field in taxi. Origin- and destination-entropy present seasonal patterns that align with the baseball season and U.S. Open. Both entropy measurements were higher from April to October, overlapping with MLB regular season and U.S. Open tournament, indicating the large geographical distribution diversity level of origins and destinations the spectators were traveling from/to.

5.2.2. Spatiotemporal change comparisons

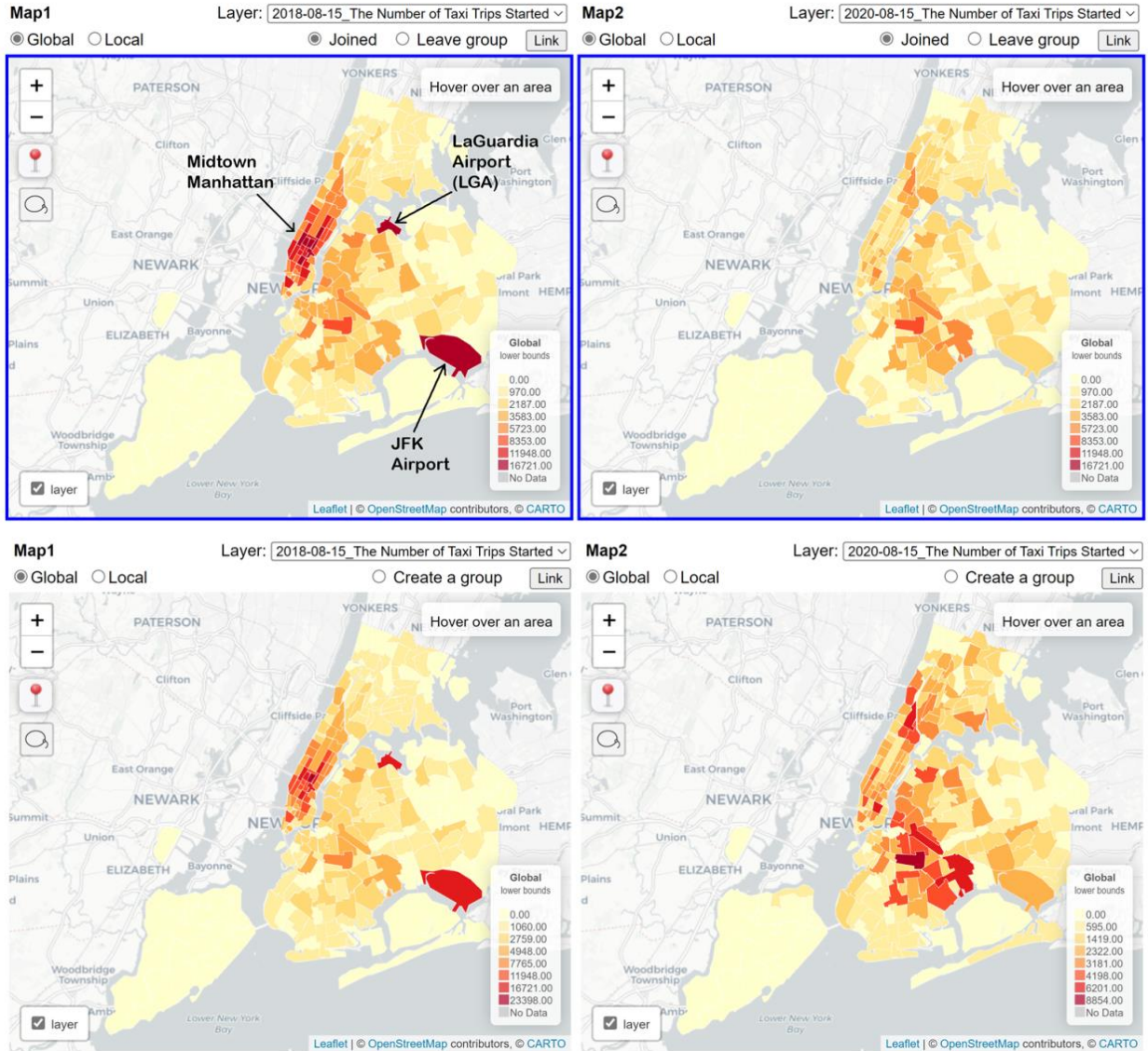

Figure 11. Origin count (the number of taxi trips started from each zone), comparing 2018-08-15 (left map) and 2020-08-15 (right map). The upper panel has consistent class intervals over two maps while the lower panel has individual class intervals for each map. Maps on the bottom panel are default visualizations. Maps having the same class intervals on the top panel can be created by setting "Link All" on the user interface.

Figure 11 presents the comparison of the number of taxi trips started from each zone (origin count) between 2018-08-15 (left two maps) and 2020-08-15 (right two maps). In the upper panel, both maps share the same class intervals enabling direct comparison, while the two maps in the lower panel have different class intervals, as shown in the legends. Comparing the two maps in the upper panel provide clear evidence of how the pandemic has significantly reduced taxi activities in NYC. The two maps in the lower panel show that the spatial distribution patterns of the taxi density were different before and during the pandemic. On 2018-08-15, the taxi zones with the largest amount of origin counts includes the two airports and midtown Manhattan. In contrast, on 2020-08-15, both airports and most of midtown Manhattan did not have many taxi trips. Instead, taxi zones with larger number of trips were mostly residential areas, located in Brooklyn, Queens, and upper Manhattan. These areas are densely populated residential areas with apartments and row-houses, where car ownership was low. After the pandemic, more residents chose taxis over public transit as their primary transportation mode, to avoid the infection. Consequently, we found that the residential areas had relatively higher origin and destination counts.

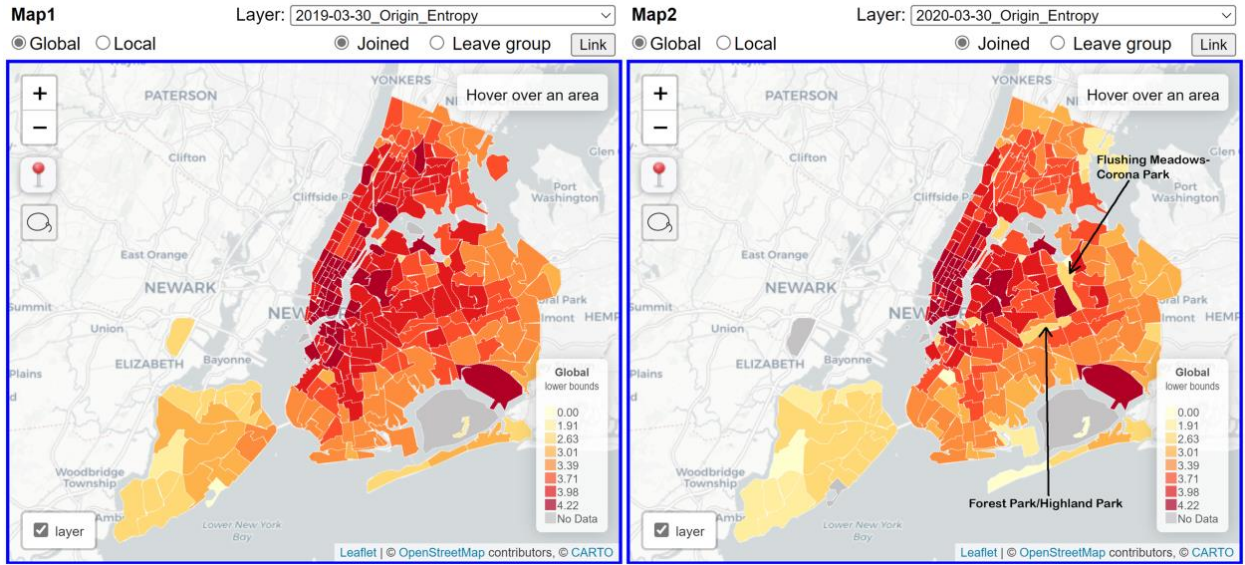

Figure 12. Origin-entropy comparison between 2019-03-30 and 2020-03-30.

Figure 12 provides an origin-entropy comparison between 2019-03-30 and 2020-03-30, with the same class intervals enabling direct comparison. These two maps show that the spatial distribution of origin entropy did not have substantial changes after the pandemic. Lower Manhattan and midtown had high origin-entropy before and during the pandemic. Despite a substantial decrease in origin count after the pandemic, midtown and lower Manhattan still had high origin-entropy. This indicates that taxi trips originated in these zones still headed to a large variety of destinations. In other words, destination distribution for people who left midtown and lower Manhattan remained similar. One possible reason for this is because of the essential positions still required in-person work, and those work-related travel patterns were less affected by the pandemic. Two taxi zones with substantial decreases in origin-entropy were Flush Meadows-Corona Park and Forest Park/Highland Park. These two zones include popular green spaces for outdoor activities and local events. During the pandemic, most events held in these places were cancelled. To elaborate further, after the pandemic, due to event cancellations and travel restrictions, people visiting these parks were more likely to do so for outdoor green spaces or essential work. Those who previously traveled there for events or tourism were unlikely to make such trips. For example, residents of Staten Island might have visited the park only for music festivals, and would not make such trips after the pandemic. Therefore, the geographical distribution diversity of where visitors were traveling to/from decreased, independent of the number of trips. This is reflected in the right map of Figure 12, which shows low entropy.

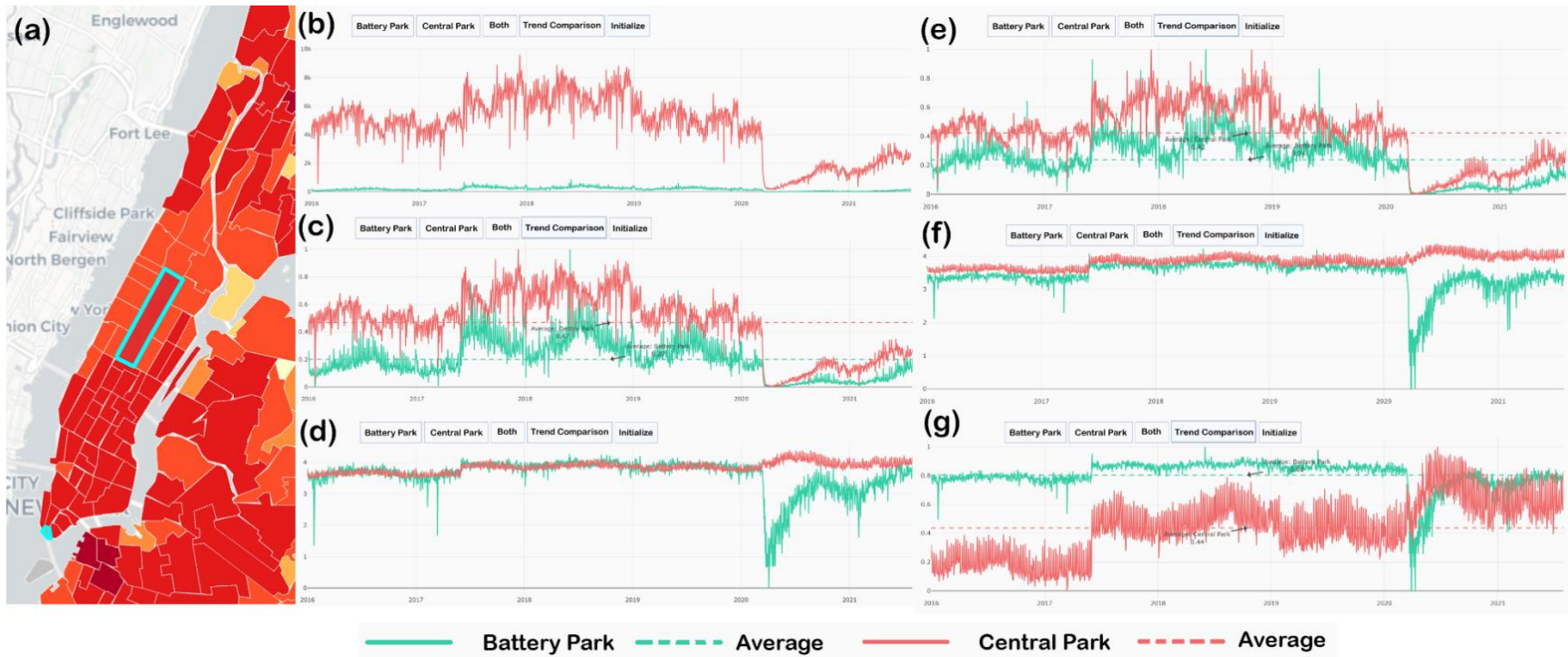

Figure 13. CLC visualizations for the comparison between Central Park and Battery Park. (a) Central Park is located in the center of Manhattan, which is highlighted by the cyan rectangle on the left map. Battery Park is at the southern tip of Manhattan which is also highlighted by cyan outline on the left map. (b) CLC visualization for origin count without normalization (under function *Initialize*). (c) CLC visualization for origin count with normalization (under function *Trend Comparison*). (d) CLC visualization for origin-entropy without normalization (under function *Initialize*). (e) CLC visualization for destination count with normalization (under function *Trend Comparison*). (f) CLC visualization for destination entropy without normalization (under function *Initialize*). (g) CLC visualization for destination-entropy with normalization (under function *Trend Comparison*). Images (b) and (c) were created from screen captures of data explored at http://su-gis.org/mobility/Taxi_CLC_Origin. Image (d) was obtained through screen captures at http://su-gis.org/mobility/Entropy_CLC_Origin. Image (e) originated from screen captures at http://su-gis.org/mobility/Taxi_CLC_Dest. Images (f) and (g) were obtained through screen captures at http://su-gis.org/mobility/Entropy_CLC_Dest.

Figure 13 shows the CLC visualization of the four measurements, providing a comparison between Central Park and Battery Park. Both parks are iconic green spaces located in Manhatton. Battery Park, located on the southern tip of Manhattan, is known for its scenic waterfront views and historical landmarks. It also includes the ferry to the Statue of Liberty, which makes it an important tourist destination in NYC. Central Park, possibly among the world's most renowned parks, is surrounded by skyscrapers in the center of Manhattan. It attracts more than 25 million New Yorkers and tourists per year. Figure 13 (a) shows the locations of Central Park and Battery Park. Although both parks are popular destinations for both local residents and tourists, the taxi-based traveling patterns are different. Figure 13 (b) compares the origin count for both parks without normalization. Obviously Central Park (red line) had a significantly larger number of trips than Battery Park (green line). After normalization, as shown in Figure 13 (c), we find that Battery Park had seasonal fluctuations, with more taxi trips being started from Battery Park in summer and less trips in winter. Central Park also presented slight seasonal fluctuation, with more taxi trips in the summer and less in winter, but it is not as dramatic as Battery Park. As there are more winter activities and indoor attractions in Central Park than in Battery Park, Central Park still attracts a large amount of visitors in the winter. In contrast, the number of visitors in Battery Park dropped substantially in winter as the temperature is low for outdoor activities. The numbers of taxi trips started from both parks were significantly dropped since the COVID-19 pandemic. Figure 13 (e) shows the normalized destination count comparison. Similar to the normalized origin count in Figure 13 (c), the normalized destination in Figure 13 (e) counts also present seasonal fluctuations in Battery Park. Destination count for both parks were dropped after the pandemic in March 2020. Figure 13 (d) provides a comparison of origin-entropy between these two parks without normalization. Figure 13 (f) and (g) show the comparison for destination-entropy without normalization (Figure 13 (f)) and with normalization (Figure 13 (g)). Although both parks are popular destinations for both tourists and locals, the entropy trend changes after the pandemic are very different. As shown in Figure 13 (d), the origin-entropy for Battery Park dropped dramatically in March 2020, as the pandemic began. On the other hand, the entropy for Central Park had a small increase since the start of the pandemic.

In other words, people who took taxi from the Central Park were traveling to a larger geographical distribution diversity of destinations compared to pre-pandemic, but people took taxi from Battery Park were traveling to a limited number of zones. The decrease in both origin- and destination-entropy of Battery Park might be due to different travel purposes, as the majority of trips were conducted by tourists or people who working there. Therefore, trips to/from Battery Park were likely traveling between a small range of origins/destinations, such as the tourism hotels and residential area where local employees live. Because of reduced tourism and visiting restrictions, both parks attracted less visitors during the pandemic, as shown in Figure 13 (c) and (e). In contrast to Battery Park, Central Park is a more preferred destination for outdoor activities among New Yorkers, even during the pandemic. In addition, people who traveled to Central Park may chose taxi over public transit to minimize the risk of infection. Therefore, we found that the both origin- and destination-entropy for Central Park increased since the pandemic. Such comparison shows the different passenger behaviors from these two parks.

## 6. Conclusion

To capture the complex human mobility patterns, this study creates an entropy-based measurement to complement the number of trips, offering a new perspective of human movements. Although the concept of entropy was widely used in measuring travel diversity and uncertainty at the individual-level, at an aggregated-level, it was more used to capture the (in)equity of distributions of geographical phenomena. We used the concept of entropy to measure the geographical distribution diversity of trips' origins and destinations. NYC taxi and ride-sharing trip records from 2016-01-01 to 2021-07-31 are used as our case study. The CyberGIS based visualizations provide opportunities to examine and compare the spatiotemporal changes for all of the four measurements.

Examining the origin count, destination count, origin-entropy, and destination-entropy of each taxi zone in NYC, we identified the changes in the geographical distribution diversity of trips' origins and destinations corresponding to major travel policy changes. After New York State legislation authorized ride-sharing services statewide in late June 2017, there was a noticeable improvement in the accessibility of ride-hailing services, including Uber, Lyft, and traditional taxicabs. This improvement was particularly evident at major hubs such as JFK Airport, where there was a significant increase in the diversity of trip origins—a concept known as higher destination-entropy—indicating a wider variety of starting points for rides to the airport than ever before. Furthermore, during the COVID-19 pandemic, the number of taxi trips sharply decreased due to strict lockdown policies in NYC. However, the diversity of trip origins and destinations increased in most areas, as evidenced by a rise in both origin-entropy and destination-entropy. This suggests a notable shift in travel behavior, with people in areas well-served by public transport increasingly turning to taxis to minimize contact with others and reduce virus transmission risks. In essence, during the pandemic, people began traveling to a more diverse range of places by taxi—locations previously frequented by public transportation. This change indicates a significant shift in transportation preferences during the pandemic.

Although the entropy-based measurements offer a unique perspective to understand the taxi-based travel patterns in the city, we realize that these measurements also have limitations. First is the limitation of data resolution. To protect privacy, NYC TLC only provides the pick-up and drop-off zone numbers, not the exact locations. In addition, the route information is also unavailable. This has limited possibilities for further analysis at a more detailed level. The second limitation is in the transportation mode choices. Although the different types of taxis were marked in the data files, we did not separate them in our measurements in this study. Conducting a more detailed measurement that separates all the taxi types (i.e., yellow taxi, green taxi, Uber, Lyft, and others) can further optimize the city regulations and dispatchments. In addition, we only considered trips done by taxi and ride-sharing services. To create a more comprehensive understanding about human flows in the city, other transportation modes, such as buses and subways, should also be considered.

The findings of this study can be used in several practical fields. First, entropy-based measurements can help future transportation optimization and urban planning. Entropy measurements can identify zones with a high geographical distribution diversity of origins or destinations, helping transportation planners to optimize infrastructure to accommodate the travel demands. Secondly, capturing the travel demand changes with tourism using entropy-based measurement can help tourism management. In the results section, we showed the seasonal fluctuations associated with sports seasons. Similarly, for tourist destinations, such as coastal areas, seasonal fluctuations are important factors for transportation planning. The entropy-based measurements can identify the changes in the geographical distribution diversity associated with tourism. Related findings can help tourism transportation management, such as increasing seasonal shuttles or algorithm improvements for high travel demands. Thirdly, in emergency responses, entropy-based measurements provide insights for evacuation and resources allocation. For example, emergency planners can use entropy-based measurements to identify major transportation hubs for supply distributions. Lastly, these entropy-based measurements can be transferred to other types of transportation origin-destination datasets in different study areas. This study, while centered on New York City using its taxi travel records as a case study, demonstrates the application of entropy measurement and CyberGIS-Vis modules that can be extended to analyze various mobility data across different regions. The replicability of this research offers significant value to researchers and policymakers aiming for a deeper comprehension of human mobility patterns across a wider geographical spectrum.

**Data availability statement**

New York City taxi data are open data. They are collected and maintained by New York City Taxi & Limousine Commission. All the data used in this study can be found and downloaded from: https://www.nyc.gov/site/tlc/about/tlc-trip-record-data.page.

Keler, A., Krisp, J. M., & Ding, L. (2017). Detecting traffic congestion propagation in urban environments–a case study with Floating Taxi Data (FTD) in Shanghai. *Journal of location Based services*, *11*(2), 133-151.

Kim, S., & Lee, S. (2023). Nonlinear relationships and interaction effects of an urban environment on crime incidence: Application of urban big data and an interpretable machine learning method. *Sustainable Cities and Society*, *91*, 104419.

Lanzendorf, M. (2002). Mobility styles and travel behavior: Application of a lifestyle approach to leisure travel. *Transportation Research Record*, *1807*(1), 163-173.

Liu, X., Gong, L., Gong, Y., & Liu, Y. (2015). Revealing travel patterns and city structure with taxi trip data. *Journal of transport Geography*, *43*, 78-90.

Liu, Y., Liu, C., Lu, X., Teng, M., Zhu, H., & Xiong, H. (2017, August). Point-of-interest demand modeling with human mobility patterns. In *Proceedings of the 23rd ACM SIGKDD international conference on knowledge discovery and data mining* (pp. 947-955).

Liu, Y., Liu, X., Gao, S., Gong, L., Kang, C., Zhi, Y., ... & Shi, L. (2015). Social sensing: A new approach to understanding our socioeconomic environments. *Annals of the Association of American Geographers*, *105*(3), 512-530.

Liu, Y., Wang, F., Xiao, Y., & Gao, S. (2012). Urban land uses and traffic 'source-sink areas': Evidence from GPS-enabled taxi data in Shanghai. *Landscape and Urban Planning*, *106*(1), 73-87.

Moncure, P. A. (2017). *A Mobility Equity Index for Evaluating Transportation Options to Access Healthy Food Outlet Points in Austin, TX, USA*. The University of Texas School of Public Health.

Naseer, K., Qazi, J., Qazi, A., Avuglah, B. K., Tahir, R., Rasheed, R. A., ... & Naseem, U. (2022). Travel behaviour prediction amid covid-19 underlaying situational awareness theory and health belief model. *Behaviour & Information Technology*, *41*(15), 3318-3328.

Ning, W., Yang, Y., Lu, M., & Han, X. (2022). Equity in walking access to community home care facility resources for elderly with different mobility: A case study of Lianhu District, Xi'an. *PLoS one*, *17*(12), e0277310.

National Weather Service. (2021, February 10). *January 31, 2021 - February 2, 2021 Winter storm.* National Weather Service, National Oceanic and Atmospheric Administration. https://www.weather.gov/okx/WinterStormJan31_Feb2202.

NY1 NEWS. (2019, October 14). 'Rolling Loud' hip-hop festival debuts to sold out crowd. *Spectrum News – NY1*. Retrieved October 26, 2023, from https://ny1.com/nyc/queens/news/2019/10/14/-rolling-loud--hip-hop-festival-debuts-to-sold-out-crowd#:~:text=Organizers%20said%2060%2C000%20fans%20sold,hosted%20in%20cities%20in%20California.

NYC TLC. (n.d.). *TLC Trip Record Data.* NYC Taxi & Limousine Commission. Retrieved October 26, 2023. https://www.nyc.gov/site/tlc/about/tlc-trip-record-data.page.

NYCEDC. (2018, April 5). *New Yorkers and Their Cars.* NYCEDC. Retrieved October 26, 2023, from https://edc.nyc/article/new-yorkers-and-their-cars.

NYU Furman Center. (n.d). *Flatbush/Midwood BK14.* New York Neighborhood Data Profiles. Retrieved October 26, 2023, from https://furmancenter.org/neighborhoods/view/flatbush-midwood.

Pan, B., Zheng, Y., Wilkie, D., & Shahabi, C. (2013, November). Crowd sensing of traffic anomalies based on human mobility and social media. In *Proceedings of the 21st ACM*
29